\catcode`\@=11

\font\tenmsa=msam10 \font\sevenmsa=msam7 \font\fivemsa=msam5
\font\tenmsb=msbm10
\font\sevenmsb=msbm7 \font\fivemsb=msbm5 \newfam\msafam \newfam\msbfam
\textfont\msafam=\tenmsa \scriptfont\msafam=\sevenmsa
\scriptscriptfont\msafam=\fivemsa \textfont\msbfam=\tenmsb
\scriptfont\msbfam=\sevenmsb \scriptscriptfont\msbfam=\fivemsb

\def\hexnumber@#1{\ifnum#1<10 \number#1\else \ifnum#1=10 A\else\ifnum#1=11
 B\else\ifnum#1=12 C\else \ifnum#1=13 D\else\ifnum#1=14 E\else\ifnum#1=15
 F\fi\fi\fi\fi\fi\fi\fi}

\def\msa@{\hexnumber@\msafam} \def\msb@{\hexnumber@\msbfam}
\mathchardef\boxdot="2\msa@00 \mathchardef\boxplus="2\msa@01
\mathchardef\boxtimes="2\msa@02 \mathchardef\square="0\msa@03
\mathchardef\blacksquare="0\msa@04 \mathchardef\centerdot="2\msa@05
\mathchardef\lozenge="0\msa@06 \mathchardef\blacklozenge="0\msa@07
\mathchardef\circlearrowright="3\msa@08 \mathchardef\circlearrowleft="3\msa@09
\mathchardef\rightleftharpoons="3\msa@0A
\mathchardef\leftrightharpoons="3\msa@0B \mathchardef\boxminus="2\msa@0C
\mathchardef\Vdash="3\msa@0D \mathchardef\Vvdash="3\msa@0E
\mathchardef\vDash="3\msa@0F \mathchardef\twoheadrightarrow="3\msa@10
\mathchardef\twoheadleftarrow="3\msa@11 \mathchardef\leftleftarrows="3\msa@12
\mathchardef\rightrightarrows="3\msa@13 \mathchardef\upuparrows="3\msa@14
\mathchardef\downdownarrows="3\msa@15 \mathchardef\upharpoonright="3\msa@16
 \mathchardef\downharpoonright="3\msa@17
\mathchardef\upharpoonleft="3\msa@18 \mathchardef\downharpoonleft="3\msa@19
\mathchardef\rightarrowtail="3\msa@1A \mathchardef\leftarrowtail="3\msa@1B
\mathchardef\leftrightarrows="3\msa@1C \mathchardef\rightleftarrows="3\msa@1D
\mathchardef\Lsh="3\msa@1E \mathchardef\Rsh="3\msa@1F
\mathchardef\rightsquigarrow="3\msa@20
\mathchardef\leftrightsquigarrow="3\msa@21 \mathchardef\looparrowleft="3\msa@22
\mathchardef\looparrowright="3\msa@23 \mathchardef\circeq="3\msa@24
\mathchardef\succsim="3\msa@25 \mathchardef\gtrsim="3\msa@26
\mathchardef\gtrapprox="3\msa@27 \mathchardef\multimap="3\msa@28
\mathchardef\therefore="3\msa@29 \mathchardef\because="3\msa@2A
\mathchardef\doteqdot="3\msa@2B 
\mathchardef\traceiangleq="3\msa@2C \mathchardef\precsim="3\msa@2D
\mathchardef\lesssim="3\msa@2E \mathchardef\lessapprox="3\msa@2F
\mathchardef\eqslantless="3\msa@30 \mathchardef\eqslantgtr="3\msa@31
\mathchardef\curlyeqprec="3\msa@32 \mathchardef\curlyeqsucc="3\msa@33
\mathchardef\preccurlyeq="3\msa@34 \mathchardef\leqq="3\msa@35
\mathchardef\leqslant="3\msa@36 \mathchardef\lessgtr="3\msa@37
\mathchardef\backprime="0\msa@38 \mathchardef\risingdotseq="3\msa@3A
\mathchardef\fallingdotseq="3\msa@3B \mathchardef\succcurlyeq="3\msa@3C
\mathchardef\geqq="3\msa@3D \mathchardef\geqslant="3\msa@3E
\mathchardef\gtrless="3\msa@3F \mathchardef\sqsubset="3\msa@40
\mathchardef\sqsupset="3\msa@41
\mathchardef\trianglelefteq="3\msa@45 \mathchardef\bigstar="0\msa@46
\mathchardef\between="3\msa@47 \mathchardef\blacktriangledown="0\msa@48
\mathchardef\blacktriangleright="3\msa@49
\mathchardef\blacktriangleleft="3\msa@4A
\mathchardef\blacktriangle="0\msa@4E \mathchardef\triangledown="0\msa@4F
\mathchardef\eqcirc="3\msa@50 \mathchardef\lesseqgtr="3\msa@51
\mathchardef\gtreqless="3\msa@52 \mathchardef\lesseqqgtr="3\msa@53
\mathchardef\gtreqqless="3\msa@54 \mathchardef\Rrightarrow="3\msa@56
\mathchardef\Lleftarrow="3\msa@57 \mathchardef\veebar="2\msa@59
\mathchardef\barwedge="2\msa@5A \mathchardef\doublebarwedge="2\msa@5B
\mathchardef\angle="0\msa@5C \mathchardef\measuredangle="0\msa@5D
\mathchardef\sphericalangle="0\msa@5E \mathchardef\varpropto="3\msa@5F
\mathchardef\smallsmile="3\msa@60 \mathchardef\smallfrown="3\msa@61
\mathchardef\Subset="3\msa@62 \mathchardef\Supset="3\msa@63
\mathchardef\Cup="2\msa@64  \mathchardef\Cap="2\msa@65
 \mathchardef\curlywedge="2\msa@66
\mathchardef\curlyvee="2\msa@67 \mathchardef\leftthreetimes="2\msa@68
\mathchardef\rightthreetimes="2\msa@69 \mathchardef\subseteqq="3\msa@6A
\mathchardef\supseteqq="3\msa@6B \mathchardef\bumpeq="3\msa@6C
\mathchardef\Bumpeq="3\msa@6D \mathchardef\lll="3\msa@6E 
\mathchardef\ggg="3\msa@6F  \mathchardef\circledS="0\msa@73
\mathchardef\pitchfork="3\msa@74 \mathchardef\dotplus="2\msa@75
\mathchardef\backsim="3\msa@76 \mathchardef\backsimeq="3\msa@77
\mathchardef\complement="0\msa@7B \mathchardef\intercal="2\msa@7C
\mathchardef\circledcirc="2\msa@7D \mathchardef\circledast="2\msa@7E
\mathchardef\circleddash="2\msa@7F \def\ulcorner{\delimiter"4\msa@70\msa@70 }
\def\urcorner{\delimiter"5\msa@71\msa@71 }
\def\llcorner{\delimiter"4\msa@78\msa@78 }
\def\lrcorner{\delimiter"5\msa@79\msa@79 } \def\yen{\mathhexbox\msa@55 }
\def\checkmark{\mathhexbox\msa@58 } \def\circledR{\mathhexbox\msa@72 }
\def\maltese{\mathhexbox\msa@7A } \mathchardef\lvertneqq="3\msb@00
\mathchardef\gvertneqq="3\msb@01 \mathchardef\nleq="3\msb@02
\mathchardef\ngeq="3\msb@03 \mathchardef\nless="3\msb@04
\mathchardef\ngtr="3\msb@05 \mathchardef\nprec="3\msb@06
\mathchardef\nsucc="3\msb@07 \mathchardef\lneqq="3\msb@08
\mathchardef\gneqq="3\msb@09 \mathchardef\nleqslant="3\msb@0A
\mathchardef\ngeqslant="3\msb@0B \mathchardef\lneq="3\msb@0C
\mathchardef\gneq="3\msb@0D \mathchardef\npreceq="3\msb@0E
\mathchardef\nsucceq="3\msb@0F \mathchardef\precnsim="3\msb@10
\mathchardef\succnsim="3\msb@11 \mathchardef\lnsim="3\msb@12
\mathchardef\gnsim="3\msb@13 \mathchardef\nleqq="3\msb@14
\mathchardef\ngeqq="3\msb@15 \mathchardef\precneqq="3\msb@16
\mathchardef\succneqq="3\msb@17 \mathchardef\precnapprox="3\msb@18
\mathchardef\succnapprox="3\msb@19 \mathchardef\lnapprox="3\msb@1A
\mathchardef\gnapprox="3\msb@1B \mathchardef\nsim="3\msb@1C
\mathchardef\napprox="3\msb@1D
\mathchardef\nsupseteqq="3\msb@23 \mathchardef\subsetneqq="3\msb@24
\mathchardef\supsetneqq="3\msb@25
\mathchardef\supsetneq="3\msb@29 \mathchardef\nsubseteq="3\msb@2A
\mathchardef\nsupseteq="3\msb@2B \mathchardef\nparallel="3\msb@2C
\mathchardef\nmid="3\msb@2D \mathchardef\nshortmid="3\msb@2E
\mathchardef\nshortparallel="3\msb@2F \mathchardef\nvdash="3\msb@30
\mathchardef\nVdash="3\msb@31 \mathchardef\nvDash="3\msb@32
\mathchardef\nVDash="3\msb@33 \mathchardef\ntrianglerighteq="3\msb@34
\mathchardef\ntrianglelefteq="3\msb@35 \mathchardef\ntriangleleft="3\msb@36
\mathchardef\ntriangleright="3\msb@37 \mathchardef\nleftarrow="3\msb@38
\mathchardef\nrightarrow="3\msb@39 \mathchardef\nLeftarrow="3\msb@3A
\mathchardef\nRightarrow="3\msb@3B \mathchardef\nLeftrightarrow="3\msb@3C
\mathchardef\nleftrightarrow="3\msb@3D \mathchardef\divideontimes="2\msb@3E
\mathchardef\varnothing="0\msb@3F \mathchardef\nexists="0\msb@40
\mathchardef\mho="0\msb@66 \mathchardef\thorn="0\msb@67
\mathchardef\beth="0\msb@69 \mathchardef\gimel="0\msb@6A
\mathchardef\daleth="0\msb@6B \mathchardef\lessdot="3\msb@6C
\mathchardef\gtrdot="3\msb@6D \mathchardef\ltimes="2\msb@6E
\mathchardef\rtimes="2\msb@6F \mathchardef\shortmid="3\msb@70
\mathchardef\shortparallel="3\msb@71 \mathchardef\smallsetminus="2\msb@72
\mathchardef\thicksim="3\msb@73 \mathchardef\thickapprox="3\msb@74
\mathchardef\approxeq="3\msb@75 \mathchardef\succapprox="3\msb@76
\mathchardef\precapprox="3\msb@77 \mathchardef\curvearrowleft="3\msb@78
\mathchardef\curvearrowright="3\msb@79 \mathchardef\digamma="0\msb@7A
\mathchardef\varkappa="0\msb@7B \mathchardef\hslash="0\msb@7D
\mathchardef\hbar="0\msb@7E \mathchardef\backepsilon="3\msb@7F
\def\Bbb{\ifmmode\let\next\Bbb@\else
\def\next{\errmessage{Use \string\Bbb\space only in math mode}}\fi\next}
\def\Bbb@#1{{\Bbb@@{#1}}} \def\Bbb@@#1{\fam\msbfam#1}

\catcode`\@=\active

 \def\CZ{\hbox{{$\cal Z$}}}

\def\R{{\Bbb R}} \def\C{{\Bbb C}}

 \def\vect{{\bf t}}\def\vecv{{\bf v}} \def\vecu{{\bf u}}\def\vecx{{\bf x}}
\def\veca{{\bf a}} \def\vecw{{\bf w}}

 \def\<{\langle} \def\>{\rangle}
\def\del{{\partial}}
\def\lform{\hbox{$\sqcup$}\llap{\hbox{$\sqcap$}}}

 \def\eps{{\epsilon}}

\def\cosub{{\Delta\kern -.65em \raisebox{.02em}{-}\kern .35em}}

 \def\tens{\mathop{\otimes}}

  \def\id{{\rm id}}
  \def\Vhaj{{V\haj{\ }}}
\def\proof{\goodbreak\noindent{\bf Proof\quad}}
 \def\endproof{{\
$\lform$}\bigskip }

\def\und#1{{\underline {#1}}} \def\haj#1{{\mathaccent20 {#1}}}

\def\Bo{{{}_{\und{\scriptscriptstyle(1)}}}}
\def\Bt{{{}_{\und{\scriptscriptstyle(2)}}}}

 \def\note#1{}

\def\equad{\kern -1.7em}  

\def\eqn#1#2{\begin{equation}#2\label{#1}\end{equation}}
\def\cmath#1{\[\begin{array}{c} #1 \end{array}\]} 
\def\ceqn#1#2{\begin{equation}\label{#1}\begin{array}{c}#2
\end{array}\end{equation}} \def\align#1{\begin{eqnarray*}#1\end{eqnarray*}}

\def\Rrel{{\bf R'}}
\def\Radd{{\bf R}}
\def\extd{{\rm d}}

\def\rdelu#1{{\overleftarrow{\del^{#1}}}}
\def\rdeld#1{{\overleftarrow{\del_{#1}}}}

\documentstyle[11pt]{article} \textheight 23.6cm \textwidth 16cm
\evensidemargin
0in \topskip 28pt

\newtheorem{lemma}{Lemma}[section] \newtheorem{propos}[lemma]{Proposition}
\newtheorem{example}[lemma]{Example} 
 \newtheorem{corol}[lemma]{Corollary}

\begin{document}\baselineskip 23pt

{\ }\hskip 4.7in DAMTP/94-66 \vspace{.2in}

\begin{center} {\LARGE $*$-STRUCTURES ON BRAIDED SPACES} \\ \baselineskip
13pt{\
} {\ }\\ S.  Majid\footnote{Royal Society University Research Fellow and Fellow
of Pembroke College, Cambridge}\\ {\ }\\ Department of Applied Mathematics \&
Theoretical Physics\\ University of Cambridge, Cambridge CB3 9EW \end{center}

\begin{center} August 1994\end{center} \vspace{10pt} \begin{quote}\baselineskip
13pt \noindent{\bf Abstract} $*$-structures on quantum and braided
spaces of the type defined via an R-matrix are studied.  These include
$q$-Minkowski and
$q$-Euclidean spaces as additive braided groups. The duality between
the $*$-braided groups of vectors and covectors is proved and some
first applications to braided geometry are made. \end{quote} \baselineskip 23pt

\section{Introduction}

The programme of q-deforming physics is an important one with possible
applications to q-regularisation as well as to models of quantum corrections to
geometry.  The first stage of this programme, to build algebras suitable for
co-ordinates of spacetime and other linear spaces (i.e.  deformations of
$\R^n)$, is fairly complete at an algebraic level.  The systematic treatment
here is the approach coming out of braided
geometry\cite{Ma:varen}\cite{Ma:introp}.  This starts with an addition law
(expressed algebraically as a braided coaddition $\und\Delta$) and proceeds
with
a systematic theory of quantum metric, Poincar\'e group\cite{Ma:poi},
differentiation\cite{Ma:fre}, integration\cite{KemMa:alg}, epsilon tensor and
differential forms\cite{Ma:eps}.  There are also natural choices in the braided
approach for $q$-deformed Euclidean\cite{Ma:euc} and
Minkowski\cite{Ma:mec}\cite{Mey:new}\cite{Mey:wav} spaces, related in by a
quantum Wick rotation.  Moreover, the braided approach is compatible with
earlier q-deformations of these particular spaces from the point of view of
$SO_q(n)$-covariance\cite{Fio:sym} and spinor
decomposition\cite{CWSSW:lor}\cite{OSWZ:def} respectively.  The relation in the
latter case is in \cite{MaMey:bra}.

Although this theory is fairly complete, it is algebraic.  For the next stage,
as well as for applications in physics, one needs to understand an abstract
approach to the $*$-structure on such spaces too.  This is needed in order to
be
able to properly formulate reality properties, Hilbert space representations
and
other key ingredients for a realisation of such non-commutative algebras in a
quantum-mechanical set-up.  A general theory of such $*$-structures has been
missing until now.

In this note we provide such a theory within the braided groups programme
mentioned above, i.e.  for deformations of $\R^n$ described by an R-matrix.  As
usual, our starting point is compatibility with the braided coaddition law.  We
give appropriate axioms for a $*$-braided group following \cite{Ma:mec} and
then
examine its compatibility with the various layers of braided geometry mentioned
above.

\subsection*{Acknowledgements}
I would like to thank J. Wess for encouragement to write up the
calculations here. Some of the results were obtained during a visit to the
 Erwin Schr\"odinger Institute in Vienna; I gratefully acknowledge their
 support.

\section{$*$-Structure on Braided Vectors and Covectors}

The most abstract definition of a $*$-braided group is not known, but one which
has already been useful in some examples is \cite{Ma:mec} that the braided
group
$B$ should be a $*$-algebra in the usual way with an antilinear
antihomomorphism
$*:B\to B$ and in addition \eqn{star}{ (*\tens
*)\circ\und\Delta=\tau\circ\und\Delta,\quad \overline{\und\eps(\
)}=\und\eps\circ*,
\quad *\circ \und S=\und S\circ *} where $(\overline{\phantom{|a}})$ denotes
complex conjugation. These are
far from the usual axioms of a Hopf $*$-algebra and so they should be since a
braided group is not a Hopf algebra in the usual sense.  Recall that for a
braided group the coproduct $\und\Delta$ is a braided-homomorphism, i.e.  an
algebra homomorphism $B\to B\und \tens B$ where the latter is the braided
tensor
product algebra \[ (a\tens c)(b\tens d)=a\Psi(c\tens b)d\] where $\Psi:B\tens
B\to B\tens B$ is a braiding\cite{Ma:exa}.  We showed in
\cite{Ma:mec} that the usual axioms of a Hopf $*$-algebra become the axioms
above under a process of transmutation that converts quantum groups into
braided
groups.  So there are some examples of (\ref{star}) known.

On the other hand, we want to apply these axioms to fresh examples not obtained
(as far as I know) by transmutation.  These are the braided vector and covector
spaces $V(R',R)=\{v^i\}$ and $\Vhaj(R',R)=\{x_i\}$ introduced in \cite{Ma:poi}.
They are defined relative to two R-matrices, with $R$ obeying the QYBE and $R'$
some mixed relations relative to $R$.  The formulae are
\eqn{covec}{\vecx_1\vecx_2=\vecx_2\vecx_1R',\quad
\Psi(\vecx_1\tens\vecx_2)=\vecx_2\tens\vecx_1 R,\quad {\rm i.e.}\quad
\vecx'_1\vecx_2=\vecx_2\vecx'_1R }
\eqn{vec}{\vecv_1\vecv_2=R'\vecv_2\vecv_1,\quad
\Psi(\vecv_1\tens\vecv_2)=R'\vecv_2\tens\vecv_1,\quad {\rm i.e.}\quad
\vecv'_1\vecv_2=R\vecv_2\vecv'_1} where the second form gives the relations of
the braided tensor product or {\em braid statistics} directly in the notation
$\vecx\equiv\vecx\tens 1,\vecx'=1\tens\vecx$ etc. With these relations,
$\vecx+\vecx'$ and $\vecv+\vecv'$ are again braided covectors and vectors
respectively. This corresponds to a linear form of $\und\Delta$ and
$\und\eps=0$
on the generators.

We will obtain various types of $*$-structure depending on the reality
properties of the matrices $R,R'$.  We consider of interest
\eqn{type1}{\overline{R^i{}_j{}^k{}_l}=R^l{}_k{}^j{}_i,\quad {\rm real\
type\ I}} \eqn{type2}{\overline{R^i{}_j{}^k{}_l}=R^{-1}{}^j{}_i{}^l{}_k,\quad
{\rm
antireal\ type\ I}} \eqn{type3}{\overline{R^i{}_j{}^k{}_l}=R^{\bar k}{}_{\bar
l}{}^{\bar i}{}_{\bar j},\quad {\rm real\ type\ II}}
\eqn{type4}{\overline{R^i{}_j{}^k{}_l}=R^{-1}{}^{\bar i}{}_{\bar j}{}^{\bar
k}{}_{\bar l},\quad {\rm antireal\ type\ II}} where the second group assumes
that our indexing set is divided into two halves related bijectively by an
involution $(\bar{\ })$ on the indices.  We use this classification for the
reality property of the matrix $R'$ as well.

\begin{lemma} If $R'$ is type II we have $*$-algebras
\cmath{
{}*:\Vhaj(R',R)\to \Vhaj(R',R),\quad x_i\mapsto x_{\bar i},\quad R'\ {\rm type\
II}\\
{}*:V(R',R)\to V(R',R),\quad v^i\mapsto v^{\bar i},\quad R'\ {\rm type\ II} }
If $R'$ is type I we have  antilinear anti-algebra isomorphisms
\cmath{*:V(R',R)\to\Vhaj(R',R),\quad v^i\mapsto x_i,\quad R'\ {\rm
type\ I}\\
{} *:\Vhaj(R',R)\to V(R',R),\quad x_i\mapsto v^i,\quad R'\ {\rm
type\ I}}
\end{lemma}
\proof We use only the reality properties of $R'$
as we do not yet consider the braiding. That $*$ as stated is an antilinear
anti-algebra
homomorphism follows at once. For example the relations $(v^iv^j)^*=
(v^bv^a)^*\overline{R'{}^i{}_a{}^j{}_b}$ requires in the real type II case
$v^{\bar j}v^{\bar i}=v^{\bar a}v^{\bar b} R'{}^{\bar j}{}_{\bar b}{}^{\bar i}
{}_{\bar a}$
which are the relations of $V(R',R)$ again. In the real type I case they
require
$x_jx_i=x_ax_b R'{}^b{}_j{}^a{}_i$, which are the
relations of $\Vhaj(R',R)$. The computation for
$*$ on the relations of $\vecx$ is similar. In the antireal cases we have
$R'{}_{21}^{-1}$
in place of $R'$ after complex conjugation, but these define the same algebra.
\endproof

Next we suppose that we have a metric $\eta_{ij}$ with transposed-inverse
$\eta^{ij}$.  It is required to obey \ceqn{metric}{ t^i{}_a
t^j{}_b\eta^{ba}=\eta^{ji}\\
\eta^{ia}\eta^{jb}R'{}^k{}_a{}^l{}_b=R'{}^i{}_a{}^j{}_b
\eta^{ak}\eta^{bl},\quad
R'{}^i{}_a{}^j{}_b \eta^{ba}=\eta^{ij}.}
The second relation here means precisely that $V(R',R)$ and $\Vhaj(R',R)$ are
isomorphic by $v^i=x_a\eta^{ai}$. This is clear and was used for example in
\cite{Mey:new}.

\begin{lemma} If $\eta$ obeys
\eqn{metric2}{ \overline{\eta^{ij}}=\eta_{ji}}
then we have $*$-algebras
\cmath{
*:\Vhaj(R',R)\to \Vhaj(R',R),\quad x_i\mapsto x_a\eta^{ai},\quad R'\ {\rm
 type\ I}\\
{}*:V(R',R)\to V(R',R),\quad v^i\mapsto  \eta_{ia} v^a,\quad R'\ {\rm  type\
I}}
\end{lemma}
\proof It is immediate from the last part of Lemma~2.1 that a real $\eta$
turns the antilinear antialgebra homomorphism there into the first map $*$ as
stated
here. We need the additional condition on $\eta$   to have $*^2=\id$. This is
compatible with the $R'$ in (\ref{metric}) for either real type I (as here) or
antireal type I. It is also compatible with equations such as (\ref{metric1})
needed below which come from covariance when $R$ is type I. Hence we have a
$*$-algebra
structure
on  the $x_i$. Likewise for the $*$-structure on braided covectors. In fact,
the
latter is such that the
isomorphism between covectors and vectors provided by the metric becomes an
isomorphism of $*$-algebras. Thus
$v^{i*}=(x_a\eta^{ai})^*=\eta_{ia}x_b\eta^{ba}=\eta_{ia}v^a$. Hence there is
really only one $*$-braided group here. Later on we will give a second
$*$-braided
group structure on braided vectors and covectors in the real type I case.
\endproof

These elementary lemmas tell us when when our algebras are $*$-algebras in the
usual sense of having an antilinear involution $*$. Note that if  $(\bar{\
})^2$
or $\eta\bar\eta$  are not the identity, we can still proceed but just lose
the condition $*^2=\id$.

We now consider when our $*$-algebras are $*$-braided groups. This depends on
the braiding $\Psi$ and is independent of the real or antireal type of
$*$-algebra.

\begin{propos} If we have a type II $*$-structure as in Lemma~2.1 and if $R$ is
real type II
then $V(R',R)$ and $\Vhaj(R',R)$ are $*$-braided groups.
\end{propos}
\proof  We have to check that the coproduct property in (\ref{star}) holds.
On quadratic elements this is
\align{\und\Delta((x_ix_j)^*)\equad&&=\und\Delta(x_j^* x_i^*)=(x_{\bar j}
\tens 1+1\tens x_{\bar j})(x_{\bar i}\tens 1+1\tens x_{\bar i})\\
&&=x_{\bar j}x_{\bar i}\tens 1+ 1\tens x_{\bar j}x_{\bar i}+x_{\bar j}
\tens x_{\bar i}
+ x_d\tens x_c R^{\bar c}{}_{\bar j}{}^{\bar d}{}_{\bar i}\\
(*\tens *)\und\Delta(x_ix_j)\equad &&=(*\tens *)(x_ix_j\tens 1+
1\tens x_ix_j+x_i\tens x_j+x_b\tens x_a R^a{}_i{}^b{}_j)\\
&&=x_{\bar j}x_{\bar i}\tens 1+ 1\tens x_{\bar j}x_{\bar i} +x_{\bar i}
\tens x_{\bar j}
+x_{\bar b} \tens x_{\bar a} R^{\bar b}{}_{\bar j}{}^{\bar a}{}_{\bar i}}
by our real type II assumption. We see that we have (\ref{star}) on quadratic
elements. We can extend this to higher products by induction. A more formal way
to see this is that $\und\Delta$ and $*$ are both extended to
products multiplicatively, where $\Vhaj\und\tens \Vhaj$ has the
braided tensor product  algebra and the $*$-structure
\[ (x_i\tens 1)^*=1\tens x_i^*,\quad (1\tens x_i)^*=x_i^*\tens 1.\]
The content of the proof above is to check that this is a well-defined
$*$-structure. Since (\ref{star}) holds on the generators, it then holds on all
products too. The proof for braided vectors is similar.

We also have to verify the other half of (\ref{star}). We have
\align{\und S((x_ix_j)^*)\equad&&=\und S(x_{\bar j}x_{\bar i})=\cdot\Psi(
\und Sx_{\bar j}\tens\und S x_{\bar i})=
=x_{\bar b}x_{\bar a} R^{\bar a}{}_{\bar j}{}^{\bar b}{}_{\bar i}\\
*\und S(x_ix_j)\equad&&=*\circ\cdot\Psi(\und Sx_i\tens\und S x_j)=(x_ax_b
R^b{}_i{}^a{}_j)^*=
x_{\bar b}x_{\bar a}\overline{R^b{}_i{}^a{}_j}}
since $\und S(\vecx)=-\vecx$ extends to products as a braided-antialgebra
homomorphism\cite{Ma:exa}.
The two expressions coincide by  the real type II assumption on $R$. The
extension
to higher powers is by induction using similar techniques. The proof for the
braided
vectors is analogous. \endproof

\begin{propos} If we have a type I $*$-structure as in Lemma~2.2 and if $R$ is
real type I
then $V(R',R)$ and $\Vhaj(R',R)$ are $*$-braided groups.
\end{propos}
\proof In the type I case we assume the real metric so that we have
$*$-algebras
from Lemma~2.2.
On quadratic elements we have
\align{\und\Delta((x_ix_j)^*)\equad&&=\und\Delta(x_j^*
x_i^*)=(x_a\eta^{aj}\tens
1+1\tens x_a\eta^{aj})(x_b\eta^{bi}\tens 1+1\tens x_b\eta^{bi})\\
&&=x_a\eta^{aj}x_b\eta^{bi}\tens 1+ 1\tens
x_a\eta^{aj}x_b\eta^{bi}+x_a\eta^{aj}\tens
x_b\eta^{bi}+\eta^{aj}\eta^{bi}x_d\tens x_c R^c{}_a{}^d{}_b\\
(*\tens *)\und\Delta(x_ix_j)\equad &&=(*\tens *)(x_ix_j\tens 1+1\tens
x_ix_j+x_i\tens x_j+x_b\tens x_a R^a{}_i{}^b{}_j)\\
&&=x_a\eta^{aj}x_b\eta^{bi}\tens 1+ 1\tens x_a\eta^{aj}x_b\eta^{bi}
+x_b\eta^{bi}\tens x_a\eta^{aj} +x_c\eta^{cb}\tens x_d\eta^{da}
R^j{}_b{}^i{}_a}
Next we use the identity
\eqn{metric1}{ \eta^{ia}\eta^{jb}R{}^k{}_a{}^l{}_b=R{}^i{}_a{}^j{}_b
\eta^{ak}\eta^{bl}}
which follows at once from the covariance condition in the first of
(\ref{metric}) by applying the fundamental R-matrix representation to
both sides. Using this and comparing our final expressions we see exactly
(\ref{star}) for the coproduct. Once again we use the braided tensor product
algebra structure
as above for a more formal proof.  The  computation for braided vectors  is
analogous.

For the other half of (\ref{star}) we have
\[\und S((x_ix_j)^*)=\und S(x_b\eta^{bj}x_a\eta^{ai})
=x_cx_d\eta^{bj}\eta^{ai}
R^d{}_b{}^c{}_a=x_cx_d\eta^{da}\eta^{cb}R^j{}_a{}^i{}_b
=(x_ax_b R^b{}_i{}^a{}_j)^*=
*\und S(x_ix_j)\]
using (\ref{metric1}) in the middle. Likewise for the braided vectors.
\endproof

A general class of examples of type II is provided by the braided matrix $B(R)$
construction in \cite{Ma:exa}. These are braided versions of $\R^{n^2}$ with
generators $\{u^i{}_j\}$ and relations $R_{21}\vecu_1
R\vecu_2=\vecu_2R_{21}\vecu_1R$.
Such relations are among the relations of quantum enveloping algebras in
 \cite{FRT:lie}, but also arose in \cite{Ma:exa} as an abstract
 quadratic algebra for braided matrices. They have a braided matrix coproduct
 $\und\Delta\vecu=\vecu\tens\vecu$ and a $*$-structure\cite{Ma:mec} making
 $B(R)$ into a $*$-bialgebra in the sense of the first half of (\ref{metric}).

  Also introduced in \cite{Ma:exa} was a multi-index notation
$u^{i_0}{}_{i_1}\equiv u_I$
 whereby the braided-matrices appeared as an $n^2$-dimensional quantum
 plane relative to a bigger matrix ${\bf R'}^I{}_J{}^K{}_L$. It was
 shown in \cite{Mey:new} that when $R$ obeys a Hecke condition  then
 there is also a braiding matrix ${\bf R}^I{}_J{}^K{}_L$ making $B(R)$ into
 braided covector space with an
 additive  $\und\Delta$ as above.  Explicitly,
 \eqn{minkR}{ \Rrel^I{}_J{}^K{}_L=R^{-1}{}^{d}{}_{k_0}{}^{j_0}{}_{a}
R^{k_1}{}_{b}{}^{a}{}_{i_0}R^{i_1}{}_c{}^b{}_{l_1} {\widetilde
R}^c{}_{j_1}{}^{l_0}{}_d,\quad \Radd^I{}_J{}^K{}_L=R^{j_0}{}_a{}^d{}_{k_0}
R^{k_1}{}_b{}^a{}_{i_0} R^{i_1}{}_c{}^b{}_{l_1} {\widetilde
R}^c{}_{j_1}{}^{l_0}{}_d.}
Meyer's additive braid statistics here can be written as
$R^{-1}\vecu_1'R\vecu_2
= \vecu_2 R_{21}\vecu_1'R$ if one wants a matrix notation.

\begin{example} If $R$ is real type I then $\Rrel$ in (\ref{minkR}) is antireal
type II and
$\Radd$ is real type II with $(i_0,i_1){}^{\bar{\ }}=(i_1,i_0)$. Hence in the
Hecke case $B(R)$ is a $*$-braided group
of the type II in Proposition~2.3. The $*$ structure is
$(u^{i_0}{}_{i_1})^*=u^{i_1}{}_{i_0}$. \end{example}
\proof The reality properties can be deduced from the form of $\Rrel,\Radd$
shown and the reality property of $R$. We then use Proposition~2.3. \endproof

The most familiar case is when $R$ is the $SU_q(2)$ R-matrix with real $q$.
Then
$B(R)$ is
q-Minkowski space and the $*$-braided group structure is the one announced in
\cite{Mey:new}\cite{MaMey:bra}. The $*$-structure   is the
Hermitian one, which is also the $*$-structure for
the multiplicative braided coproduct introduced in \cite{Ma:mec}. So $B(R)$ is
a
$*$-braided group for both coaddition and comultiplication simultaneously.

Closely related to $B(R)$ (by a cocycle twisting) is the algebra $\bar A(R)$
in \cite{Ma:euc}. This also has a matrix $x^{i_0}{}_{i_1}=x_I$ of generators
and in the Hecke case also forms a braided covector space with\cite{Ma:euc}
\eqn{eucR}{
 {\bf R'}^I{}_J{}^K{}_L=R^{-1}{}^{l_0}{}_{k_0}{}^{j_0}{}_{i_0}
R^{i_1}{}_{j_1}{}^{k_1}{}_{l_1},\quad
 {\bf
R}^I{}_J{}^K{}_L=R{}^{j_0}{}_{i_0}{}^{l_0}{}_{k_0}
R^{i_1}{}_{j_1}{}^{k_1}{}_{l_1}}
The corresponding relations and additive braid statistics can be written as
$R_{21}\vecx_1\vecx_2=\vecx_2\vecx_1R$ and $\vecx'_1\vecx_2=R\vecx_2\vecx'_1R$
if one wants a matrix notation.

\begin{example} If $R$ is real type I then $\Rrel$ and $\Radd$ in
(\ref{eucR}) are real type I.
Hence if $R$ is Hecke and if there is an invariant metric $\eta^{IJ}$ obeying
(\ref{metric2}) then
 $\bar A(R)$ is a $*$-braided group
of the type I in Proposition~2.4. The $*$-structure is
$(x^{j_0}{}_{j_1})^*=x^{i_0}{}_{i_1}\eta^{IJ}$. \end{example}
\proof The reality properties are immediate from the form of $\Rrel,\Radd$
shown. We then use Proposition~2.4. \endproof

The most familiar case is again when $R$ is the $SU_q(2)$ R-matrix with real
$q$.  Then $\bar A(R)$ is q-Euclidean space and the $*$-braided group structure
is
\[ \pmatrix{a^*&b^*\cr c^*&d^*}=\pmatrix{d&-q^{-1}c\cr -qb&a}\]
as found another way in \cite{Ma:euc}. We use the construction in Example~2.6
with
\[ \eta^{IJ}=\pmatrix{0&0&0&1\cr 0&0&-q&0\cr 0&-q^{-1}&0&0\cr 1&0&0&0}.\]
It is evident that $\eta^2=\id$ as required for (\ref{metric2}) when $\eta$
is real.

For the specific $SU_q(2)$ R-matrix the above $*$-structures on $q$-Minkowski
and $q$-Euclidean
space agree with those used in other more specific approaches
\cite{OSWZ:def} and
\cite{Fio:sym} respectively. On the other hand, the $*$ properties of
the braided coaddition have not been considered in any detail before. Moreover,
our
analysis is quite general and applies just as well to other real type
R-matrices,
including non-standard and multiparameter ones.

Our analysis also gives something more unexpected in the type I case. The
equations (\ref{metric})--(\ref{metric1}) which we used are invariant under
$\eta^{ij}$ replaced by $\eta^{ji}$. Since these were all that we really
needed to check the $*$-braided group structure, we have:

\begin{corol} When $R'$ is type $I$ and $R$ real typ I with a metric obeying
(\ref{metric2}), we have second $*$-structures
\cmath{
\star:\Vhaj(R',R)\to \Vhaj(R',R),\quad x_i\mapsto \eta^{ia} x_a \\
\star:V(R',R)\to V(R',R),\quad v^i\mapsto  v^a\eta_{ai} }
also making the braided vectors and covectors into $*$-braided groups.
\end{corol}

For example, $q$-Euclidean space has a second structure
\[ \pmatrix{a^\star&b^\star\cr c^\star&d^\star}=\pmatrix{d&-qc\cr
-q^{-1}b&a}.\]
One can check that the first $*$-structure from Lemma~2.2 is such that the
usual right coaction of the background
quantum group $\vect$ is a $*$-algebra homomorphism  when we take the usual
$*$-structure $t^i{}_j{}^*=St^j{}_i$ appropriate for $R$ of real type I. On the
other hand, this second $\star$ operation does not have this property.

\section{Duality Pairing under $*$}

We define the duality pairing of vectors and covectors by the braided binomials
\ceqn{vxpairing}{\<v^{j_m}\cdots
v^{j_2}v^{j_1},x_{i_1}x_{i_2}\cdots
x_{i_r}\>=\delta_{m,r}([m;R]!)^{j_1j_2\cdots
j_m}_{i_1i_2\cdots i_m}}
where
\cmath{{} [m;R]!=[2;R]_{m-1\,
m}[3;R]_{m-2\, m}\cdots [m;R]_{1\cdots m}\\
{}[m;R]=1+(PR)_{12}+(PR)_{12}(PR)_{23}+\cdots (PR)_{12}\cdots (PR)_{m-1\, m} }
are the braided factorial and braided integer matrices introduced in
\cite{Ma:fre}. The numerical suffices refer as usual to the copies of a tensor
product
of matrices.
This pairing is an immediate application of the braided-differentiation
in the $\vecx$ co-ordinates in
\cite{Ma:fre}. See also \cite{KemMa:alg}. We will need to consider also
\cmath{[m;R]^{\rm op}!\equiv [m;R]_{1\cdots m}^{\rm op}\cdots
[3;R]_{m-2\, m}^{\rm op}[2;R]_{m-1\, m}^{\rm op}\\
{}[m;R]^{\rm
op}=1+(PR)_{12}+(PR)_{23}(PR)_{12}+\cdots+(PR)_{m-1\, m}\cdots (PR)_{12}}
which is how the pairing comes out if we differentiate in the $\vecv$
co-ordinates
instead.

\begin{lemma} For every matrix $R$ obeying the QYBE we have the identities
\cmath{
[m;R]^{\rm op}!=[m;R]!
=([m;R_{21}]!)_{m\cdots 21}.}
\end{lemma}
\proof We use the QYBE or braid relations
$(PR)_{12}(PR)_{23}(PR)_{12}
=(PR)_{23}(PR)_{12}(PR)_{23}$ repeatedly. At order three we have
\align{[3;R]^{\rm op}!\equad &&=(1+(PR)_{12}+(PR)_{23}(PR)_{12})(1+(PR)_{23})\\
&&=1+(PR)_{12}+(PR)_{23}(PR)_{12}+
(PR)_{23}+(PR)_{12}(PR)_{23}+(PR)_{23}(PR)_{12}(PR)_{23}\\
{}[3;R]!\equad &&=(1+(PR)_{23})(1+(PR)_{12}+(PR)_{12}(PR)_{23})\\
&&=1+(PR)_{12}+(PR)_{12}(PR)_{23}+(PR)_{23}+(PR)_{23}(PR)_{12}+(PR)_{23}
(PR)_{12}(PR)_{23}}
which holds automatically, while higher $m$ require the braid relations.
Likewise
\align{[3;R_{21}]!_{321}\equad
&&=\left( (1+(RP)_{23})(1+(RP)_{12}+(RP)_{12}(RP)_{23})\right)_{321}\\
&&=(1+(RP)_{21})(1+(RP)_{32}+(RP)_{32}(RP)_{21})\\
&&=(1+(PR)_{12})(1+(PR)_{23}+(PR)_{23}(PR)_{12})\\
&&=1+(PR)_{23}+(PR)_{23}(PR)_{12}+(PR)_{12}+(PR)_{12}(PR)_{23}+(PR)_{12}
(PR)_{23}(PR)_{12}}
which coincides with  $[3;R]!$ above by the braid relations. Similarly for
higher $m$. The formal proof is best done diagramatically. On the other hand,
we will give more abstract reasons from the point of braided groups why these
non-trivial identities hold. \endproof

Next, we say that two $*$-braided groups are in duality if there is a pairing
\eqn{pair*}{\overline{\<b,c\>}=\<b^*,c^*\>}
These axioms appear quite different from the usual ones for Hopf $*$-algebras,
in particular not involving the antipode. But they are the appropriate ones
if we want consistency with our definition (\ref{metric}) and the definition of
duality for the  braided group structure. As explained in \cite{Ma:introp} it
is
convenient (but not essential) to define this as
\eqn{pairing}{ \<ab,c\>=\<a,\<b,c\Bo\> c\Bt\>,\quad \<a,cd\>=\<a\Bo,\<a\Bt,c\>
d\>,\quad
\<\und Sa,c\>=\<a,\und S c\> }
where $\und\Delta c=c\Bo\tens c\Bt$ etc. This is not the usual pairing because
we do not move $b$ past $c\Bo$ to evaluate on $c\Bt$ etc, as one would
usually do. It is possible to define such a more usual pairing by using the
braiding $\Psi$
to make the transposition but the result would be equivalent to (\ref{pairing})
via
the
braided antipode, so we avoid such an unnecessary complication. For braided
groups
paired in this way, (\ref{pair*}) is the correct extension to the $*$ case,
as one may easily check.

\begin{propos} The $*$-braided groups $V(R',R)$ and $\Vhaj(R',R)$ of type II
from
Proposition~2.3  are in duality.
\end{propos}
\proof We consider the  setting with $R$ real type II. Then
\align{\overline{\<v^{j_m}\cdots v^{j_1},x_{i_1}\cdots x_{i_m}\>}\equad &&=
\overline{([m;R]!)^{j_1\cdots j_m}_{i_1\cdots
i_m}}=([m;R_{21}]!)^{\bar{j}_1\cdots \bar{j}_m}_{\bar{i}_1\cdots
\bar{i}_m}=([m;R]!)^{\bar{j}_m\cdots \bar{j}_1}_{\bar{i}_m\cdots \bar{i}_1}
\\&&=\<v^{\bar{j}_1}\cdots v^{\bar{j}_m},x_{\bar{i}_m}\cdots x_{\bar{i}_1}\>=
\<(v^{j_m}\cdots v^{j_1})^*,(x_{i_1}\cdots x_{i_m})^*\>}
as required. We used the first part of Lemma~3.1 and $*$ from the
first part of Lemma~2.1. \endproof

\begin{propos} The $*$-braided group $V(R',R)$ of type I from Corollary~2.7
and
$\Vhaj(R',R)$ of type I from Proposition~2.4 are in duality.
\end{propos}
\proof We need here the braided vectors with the second $\star$-structure
mentioned in
Section~2. The same result applies  if we take braided covectors with
their second $\star$ operation and braided vectors with their  original one
from Proposition~2.4.

We note first that $R$ real type I means that
$\overline{(PR)^i{}_j{}^k{}_l}=(PR)^j{}_i{}^l{}_k$ or
$\overline{PR}=(PR)^{t\tens
t}$ where $t$ denotes transposition in the matrix indices. Since transposition
also reverses the order of matrix multiplication we see that
\[ \overline{[m;R]}=([m;R]^{\rm op})^{t\tens\cdots\tens t},\quad
\overline{[m;R]!}=([m;R]^{\rm op}!)^{t\tens\cdots\tens t}.\]
We also suppose a suitable metric $\eta$ as above and deduce from
repeated applications of (\ref{metric1}) that
\eqn{braintmet}{ [m;R]^{a_1\cdots a_m}_{b_1\cdots b_m}\eta^{b_1i_1}\cdots
\eta^{b_mi_m}=
\eta^{a_1b_1}\cdots\eta^{a_mb_m}([m;R_{21}]^{\rm op})^{i_1\cdots
i_m}_{b_1\cdots
b_m}}
{}From this we deduce further that
\[([m;R]!)^{a_1\cdots a_m}_{b_1\cdots b_m}\eta^{b_1i_1}\cdots
\eta^{b_mi_m}=
\eta^{a_1b_1}\cdots\eta^{a_mb_m}([m;R_{21}]^{\rm op}!)^{i_1\cdots
i_m}_{b_1\cdots b_m}=\eta^{a_1b_1}\cdots\eta^{a_mb_m}([m;R]!)^{i_m\cdots
i_1}_{b_m\cdots b_1}\]
using (\ref{braintmet}) repeatedly for the first equality and Lemma~3.1 for the
second.

We can now compute
\align{&&\equad \overline{\<v^{j_m}\cdots v^{j_1},x_{i_1}\cdots
x_{i_m}\>}=\overline{([m;R]!)^{j_1\cdots j_m}_{i_1\cdots
i_m}}\\
&&=([m;R]^{\rm op}!)^{i_1\cdots i_m}_{j_1\cdots j_m}=([m;R]!)^{i_1\cdots
i_m}_{j_1\cdots j_m}= ([m;R]!)^{a_m\cdots
a_1}_{b_m\cdots
b_1}\eta_{a_1j_1}\cdots\eta_{a_mj_m}\eta^{b_mi_m}\cdots\eta^{b_1i_1}
\\&&=\<v^{a_1}\cdots v^{a_m},x_{b_m}\cdots
x_{b_1}\>\eta_{a_1j_1}\cdots\eta_{a_mj_m}\eta^{b_mi_m}\cdots\eta^{b_1i_1}=
\<(v^{j_m}\cdots v^{j_1})^\star,(x_{i_1}\cdots x_{i_m})^*\>}
using  the above observations and the definition of $*$ from Lemma~2.2 and
$\star$
from
Corollary~2.7.
\endproof

This is the abstract reason that the identities in Lemma~3.1 must hold,
i.e. one can  essentially push these arguments backwards using (\ref{pairing})
and (\ref{pair*}) to define the dual $*$-braided group to the braided
covectors,
and computing it on the generators as the braided vectors. Such an argument
makes
sense in nice cases where the pairing is non-degenerate, which we have not
assumed
in our direct treatment above.

Finally, we mention what can be done in the case when $R$ is of type I but
there
is no invariant metric $\eta$ as we needed for a $*$-braided group structure in
Proposition~2.4. In this case we use the maps $*$ in the second part of
Lemma~2.1
not as a $*$-structure on one algebra but as a map between the pair consisting
of
the vectors and covectors. This leads to the notion of what could be
called a {\em braided holomorphic structure}. By definition we consider this to
consist of $(B,C,\<\ ,\ \>,*)$ where $B,C$ are two braided groups, $\<\ ,\
\>:B\tens C\to \C$ is a duality pairing between them and  $*$ denotes
a mutually inverse pair of antilinear antialgebra
homomorphisms $*:B\to C$ and $*:C\to B$ such that
(\ref{star}) holds between the coproducts etc. of $B,C$ and (\ref{pair*}) holds
in the form
\eqn{holpair}{ \overline{\<b,c\>}=\<c^*,b^*\>.}
We see by the same techniques as above that if $R'$ is type I and $R$ is real
type I then $V(R',R),\Vhaj(R',R)$ have such a holomorphic structure. Indeed, we
have
\align{ \overline{\<v^{j_m}\cdots v^{j_1},x_{i_1}\cdots
x_{i_m}\>}\equad&&=\overline{([m;R]!)^{j_1\cdots j_m}_{i_1\cdots
i_m}}=([m;R]^{\rm op}!)^{i_1\cdots i_m}_{j_1\cdots j_m}=([m;R]!)^{i_1\cdots
i_m}_{j_1\cdots j_m}\\
&&=\<v^{i_m}\cdots v^{i_1},x_{j_1}\cdots x_{j_m}\>=\<(x_{i_1}\cdots
x_{i_m})^*,(v^{j_m}\cdots v^{j_1})^*\>}
without requiring a metric. This covers the important case of the quantum
planes
of $GL_q(n)$ type
(such as the usual 2-dimensional quantum plane $yx=qxy$). They are known to be
braided groups\cite{Ma:poi}.

\section{$*$-Properties of Differentials and Related Structures}

As a first application of the above duality theory of $*$-braided groups, we
consider braided differentiation on braided covectors $x_i$. This was
introduced
in \cite{Ma:fre} as an infinitesimal coproduct,
\[ \del^if(\vecx)={\rm coeff}\ a_i\ {\rm in} f(\veca+\vecx)\]
which, using the braided-binomial theorem\cite{Ma:fre} gives
\[ \del^ix_{i_1}\cdots x_{i_m}=x_{j_2}\cdots x_{j_m}[m;R]^{ij_2\cdots
j_m}_{i_1\cdots i_m}.\]
These operators obey the relations of the vector algebra $V(R',R)$ and a
braided-Leibniz rule which can written in the quantum-mechanical form
\eqn{lweyl}{\del^i x_j-x_a R^a{}_j{}^i{}_b\del^b=\delta^i{}_j}
as studied by several authors. Here $x_i$ are multiplication operators from the
left.
We will also need the right-handed derivatives
\cmath{f(\vecx)\rdelu  i= f(\vecx+\veca)|_{{\rm coeff}\ a_i},\qquad
 x_{i_m}\cdots x_{i_1}\rdelu  i=x_{j_m}\cdots x_{j_2}[m;R_{21}]^{ij_2\cdots
j_m}_{i_1\cdots i_m}}
 obtained from the binomial coefficient matrix $\left[{m\atop
m-1};R\right]=[m;R_{21}]_{m\cdots 21}$ also given in \cite{Ma:fre}.
The $\rdelu  i$ obey the relations of $V(R',R)$ when considered acting this
time
on
$f(\vecx)$ from the right. They obey a right-handed braided-Leibniz rule. In
terms
of operators $x_i$ of multiplication from the right, the latter is
\eqn{rweyl}{x_j\rdelu i-\rdelu b R^i{}_b{}^a{}_j x_a =\delta^i{}_j.}

These differentiations can be defined in terms of the pairing of vectors
and covectors. Thus
\[ \del^if=\<v^i,f\Bo\>f\Bt,\quad f\rdelu  i=f\Bo\<v^i,f\Bt\>\]
from which their abstract properties (such as the assertions that they
are left and right actions of the vector algebra) follow easily. On the other
hand, since we know from Section~3 how the pairing behaves under $*$, we deduce
at
once from Proposition~3.2 or~3.3 that
\eqn{del*}{(\del^i f)^*=\overline{\<v^i,f\Bo\>}(f\Bt)^*
=\<v^{i*},f^*\Bt\>f^*\Bo=f^* \rdelu  {i*}=\cases{f^* \rdelu  a\eta_{ai}&\
type\
I\cr f^* \rdelu  {\bar i}&\ type\ II.}}
We just used (\ref{pair*}) and (\ref{metric}). One can verify this directly
also using the reality property of $R$ in the type I,II cases and
(\ref{braintmet}) in the first case. Likewise,
\eqn{6*}{(f\rdelu
i)^*=(f\Bo)^*\overline{\<v^i,f\Bt\>}=f^*\Bt\<v^{i*},f^*\Bo\>=\del^{i*}f^*=
\cases{\eta_{ai}\del^af^*&\ type\ I\cr
\del^{\bar i}f^*&\ type\ II.}}

If we are in the holomorphic setting where $R$ is type I but without using a
metric, we need a different concept, namely that of differentiation (from the
right or left) in the $v^i$ co-ordinates. From the right this is
\cmath{f(\vecv)\rdeld i= f(\vecv+\vecw)|_{{\rm coeff}\ w_i},\qquad
 v^{i_m}\cdots v^{i_1}\rdeld i=v^{j_m}\cdots v^{j_2}([m;R]^{\rm
op}){}_{ij_2\cdots
j_m}^{i_1\cdots i_m}.}
These $\rdeld i$ obey the relations of the braided covector algebra when acting
from the
right, and
\eqn{holdel*}{
 (\del^if(\vecx))^*=\overline{\<v^i,f\Bo\>}(f\Bt)^*
=f^*\Bo\<f^*\Bt,x_i\>=f(\vecx)^*
\rdeld i}
in the holomorphic case. The $GL_q(n)$ R-matrix provides an example and
is compatible with the holomorphic  calculus in \cite{Kem:sym} for this
example.

Next, the braided exponential $\exp(\vecx|\vecv)$ can be defined abstractly as
coevaluation\cite{KemMa:alg}, cf.\cite{Ma:fre}. By this we mean the
canonical element in $\Vhaj(R',R)\tens V(R',R)$ albeit as a formal powerseries
since these algebras are not finite-dimensional. This makes sense generically,
when the pairing between braided vectors and covectors is non-degenerate so
that we can in principle come up with a basis and dual basis. The exact form
of the
braided exponential depends on the algebras. Nevertheless, without knowing its
exact form we deduce from Proposition~3.2 or~3.3 that
\eqn{exp*}{ (*\tens *)\exp(\vecx|\vecv)=\exp(\vecx|\vecv)}
whenever $\exp$ can be determined as the canonical element (or coevaluation)
for the pairing. In the holomorphic case we have $\tau(\exp(\vecx|\vecv))$
instead on the right hand side.

In differential terms we can also characterise $\exp$ as the solution of
\[
\del^i\exp(\vecx|\vecv)=\exp(\vecx|\vecv)v^i,\quad
\exp(\vecx|\vecv)\rdeld i=x_i\exp(\vecx|\vecv)\]
with the usual conditions at $0$ expressed via $\und\eps$. These equations
are
equally well solved by $(*\tens *)\exp(\vecx|\vecv)$  or its transpose in the
holomorphic case.

For example, if $R'=P$ we can develop $\exp$ as a formal powerseries with terms
of the form $\vecx_1\cdots\vecx_m([m;R]!)^{-1}\vecv_m\cdots\vecv_1$ in a
compact
notation. In this case we see (\ref{exp*}) etc. immediately from Lemma~3.1. On
the other hand, our analysis is not tied to this special case.

We have also studied braided Gaussians in \cite{KemMa:alg} as the solution of
the equation
\[ \del^ig=-x_a \eta^{ai} g\]
where a metric is supposed. Applying $*$ to this gives the equation
\eqn{gauss*}{g^*\rdelu  i=-g^*\eta^{ia} x_a}
provided
\eqn{metric3}{ \overline{\eta^{ij}}=\cases{\eta_{ji}&\ type\ I\cr
\eta^{\bar j\, \bar i}&\ type\ II.}}
This is a natural condition in the real type II case since the various R-matrix
identities obeyed by $\eta$ are then mapped consistently by complex
conjugation.
In the
real type I case we already demanded the above constraint and for the same
reasons. These conditions on the metric are also appropriate when $R',R$ are
antireal except
that $R$ should be antireal when computed in the quantum group normalisation.

If $R,\eta$ obey some additional conditions as explained in \cite{KemMa:alg}
then
one
has an explicit form for $g$ as a q-exponential of
$\vecx\cdot\vecx=x_ax_b\eta^{ba}$. This obeys
$(\vecx\cdot\vecx)^*=\vecx\cdot\vecx$ in either type I, type II case when we
have
the
conditions (\ref{metric3}) on $\eta$. In these cases, which include
$q$-Minkowski and
$q$-Euclidean spaces we have $g^*=g$ for real $q$. The Euclidean space example
agrees
with more specific computations   in \cite{Fio:sym}.

Next, translation-invariant integration $\int$ was studied in \cite{KemMa:alg}
in terms of the Gaussian weighted average $\CZ[f(\vecx)]=(\int f(\vecx)g)(\int
g)^{-1}$, which we gave using (\ref{lweyl}) as
\cmath{ \CZ[1]=1,\quad \CZ[x_i]=0,\quad \CZ[x_ix_j]=\lambda^2 \eta_{ba}
R^a{}_i{}_b{}_j\\
\CZ[x_{i_1}\cdots x_{i_m}]=\sum_{r=0}^{m-2}\CZ[x_{i_1}\cdots
x_{i_r}x_{a_{r+3}}\cdots x_{a_m}]\CZ[x_{i_{r+1}}x_{a_{r+2}}]
[r+2,m;R]^{a_{r+2}\cdots a_m}_{i_{r+2}\cdots i_m}\lambda^{2(m-2-r)}}
where $\lambda$ is the quantum group normalisation constant and
$[1,m;R]=(PR)_{12}\cdots (PR)_{m-1\, m}$
in the notation of \cite{Ma:fre}. On the other hand, there are in principle
two translation-invariant integrals according to
whether we use the coaddition $\und\Delta$ from the left or the right. In
differential
terms it is natural to characterise them relative to $g$ and $g^*$ (say) as
\[ \int \del^i (f(\vecx)g)=0,\quad \int_R(g^* f(\vecx))\rdelu i=0\]
for polynomials $f$. An analogous calculation to that in \cite{KemMa:alg} but
now using
(\ref{gauss*}) and (\ref{rweyl}) gives the ratio for $\int_R g^*f(\vecx)$
against $\int_R g^*$ as
\[ \CZ_R[x_{i_1}\cdots x_{i_m}]=\sum_{r=0}^{m-2}\CZ_R[x_{a_1}\cdots
x_{a_r}x_{i_{r+3}}\cdots x_{i_m}]\CZ[x_{a_{r+1}}x_{i_{r+2}}] ([1,r+1;R]^{\rm
op})^{a_1\cdots a_{r+1}}_{i_1\cdots i_{r+1}}\lambda^{2r}\]
where $[1,m;R]^{\rm op}=(PR)_{m-1\, m}\cdots (PR)_{12}$. Using these explicit
formulae one can see by the
same techniques as in the proofs of Proposition~3.2 and~3.3 above that
\eqn{zint*}{ \CZ[(x_{i_1}\cdots x_{i_m})^*]=\overline{\CZ_R[x_{i_1}\cdots
x_{i_m}]}}
for the real type I or II cases when $\lambda$ is real. To see this note that
this is true for
$\CZ[x_ix_j]=\CZ_R[x_ix_j]$ by the assumptions (\ref{metric3}). We then proceed
by induction, using the
identities
\[ \overline{[1,m;R]^{\rm op}}=[1,m;R_{21}]^{\rm op}=[1,m;R]_{m\cdots 21}\]
in the real type II case (with barred indices on the right). In the real type I
case we use
\[ [1,m;R]^{a_1\cdots a_m}_{b_1\cdots b_m}\eta^{b_1i_1}\cdots
\eta^{b_mi_m}=\eta^{a_1b_1}\cdots\eta^{a_mb_m}([1,m;R_{21}]^{\rm
op})^{i_1\cdots i_m}_{b_1\cdots
b_m},\quad \overline{[1,m;R]}=([1,m;R]^{\rm op})^{t\tens\cdots\tens t}.\]
Note that if $g^*=g$ is central and $\int=\int_R$ (the unimodular case) then we
can conclude from
(\ref{zint*}) that the latter is complex-linear as we might expect. On the
other hand, let us
stress that the ratios $\CZ,\CZ_R$ can be used even when the integrals and
Gaussians themselves have no
meaning.

Next, we have explained in \cite{Ma:eps} the braided approach
to the calculus of differential forms and $\eps$ tensors on a braided space.
Supposing also that $R'$ obeys the QYBE and other identities, we formally
reverse the roles $R\leftrightarrow -R'$. The braided covector space
$\Vhaj(-R,-R')$ is the algebra of forms with generators $\theta_i=\extd
x_i$. Hence if $R$ is real type I (with a metric) or real type II, we have a
$*$-algebra of forms by Lemma~2.2 or~2.1 respectively. We conclude that the
action on forms $\theta_i$ is the same as on $x_i$, i.e. $(\extd x_i)^*=\extd
(x_i)^*$. The main difference however, between the braided approach and the
usual approach to forms based on the axioms of
Woronowicz\cite{Wor:dif}\cite{WesZum:cov} is that in the braided approach
the operator $\extd$ is constructed on the algebra generated by $\theta_i,x_i$
rather than being posited axiomatically. We have
\[ \left(\theta_{i_1}\cdots\theta_{i_m} f(\vecx)\right)
\overleftarrow\extd=\theta_{i_1}\cdots\theta_{i_m} \theta_i\del^if(\vecx),\quad
\vecx_1\theta_2=\theta_2\vecx_1R.\]
The exterior algebra here is the braided tensor product $\Omega_R$ of forms and
co-ordinates. On the other hand, we can equally well define $\Omega_L$ as the
braided tensor product of co-ordinates and forms instead, and
\[ \extd
\left(f(\vecx)\theta_{i_1}\cdots\theta_{i_m}\right)=f(\vecx)\rdelu
i\theta_i\theta_{i_1}\cdots
\theta_{i_m},\quad \theta_1\vecx_2=\vecx_2\theta_1R\]
instead. The operator $\overleftarrow{d}$ is the right-handed exterior
derivative in \cite{Ma:eps} and is a right-handed
super-derivation on $\Omega_R$. By reflecting the diagram-proofs there and
reversing braid-crossings, or by
explicit calculation from (\ref{rweyl}) one sees that $\extd$ here is a more
usual left-derivation on $\Omega_L$.

On the other hand,  neither of our left or right exterior algebras are
naturally $*$-algebras in general. Instead, as we noted in the
proof of Proposition~2.3 and \cite{Ma:mec}, the natural operation takes us
from one braided tensor product algebra to the reversed one:
\[ *:\Omega_R\to\Omega_L,\quad (\theta_i\tens 1)^*=1\tens \theta_i^*,\quad
(1\tens x_i)^*=x_i^*\tens 1\]
and similarly for $*$ in the other direction. Then we have
\eqn{extd*}{ (\omega\overleftarrow{\extd})^*=\extd(\omega^*)}
for any $\omega\in \Omega_R$. Note that $x_i\overleftarrow{\extd}=\extd
x_i=\theta_i$.

Finally, we defined upper and lower $\eps$ tensors in \cite{Ma:eps} by
differentiation of the
top form in form-space, leading to general formulae for them in terms of
braided-factorials $[n;-R']!$. From Lemma~3.1 we conclude at once that
if $R'$ is of real type I then
\eqn{eps*}{\overline{\eps_{i_1i_2\cdots i_n} }=\eps^{i_n\cdots i_2i_1}}
in the  construction of \cite{Ma:eps}.

\section{Concluding Remarks}

There are several open problems in $q$-deformed physics which deserve
serious attention and where an abstract understanding of the $*$-structure
along the lines of the above should be useful. Probably the most important
is to find an appropriate $*$-structure on the Weyl (or quantum
mechanics) algebra (\ref{lweyl}). Likewise to find an
appropriate $*$-structure on the Poincar\'e Hopf algebra generated by
$\del^i$ and $q$-Lorentz transformations.

The Weyl algebra here was studied for the $GL_q(n)$ R-matrices in
\cite{PusWor:twi}\cite{Kem:sym} and elsewhere, while the general R-matrix
or braided point of view was given in \cite{Ma:fre}. In the latter we showed
that
this algebra is abstractly a braided semidirect product of the braided
group of vectors acting on the braided covectors. Since we have obtained
natural
$*$-structures on these objects above, one can expect that there is a suitable
theory of semidirect products by $*$-braided groups to apply here. It is hoped
to present such a theory elsewhere.

The Poincar\'e algebra was studied in the Minkowski case by hand in
\cite{OSWZ:def}, with a general braided groups construction giving the
Minkowski, Euclidean and other Poincar\'e group quantum function algebras, due
to
the author in \cite{Ma:poi}. The general construction is again by means of an
abstract semidirect product procedure
(bosonisation) applied to the braided group of covectors.  The above
$*$-braided
group structure on the braided
covectors, and a standard Hopf $*$-algebra structure on the rotations or
Lorentz
sector imply together a $*$-structure for the resulting Poincar\'e quantum
group. But because it is a hybrid object built partly from a braided group and
partly from a quantum one (for the rotations etc.) one need not expect it to be
a usual Hopf $*$-algebra. Instead it
obeys some hybrid axioms to be elaborated elsewhere.

Once the appropriate $*$-structure and its properties are known in these
two
cases, one can proceed to the consideration of Hilbert space representations of
these structures. These are some directions for further work.

\end{document}